\documentclass[12pt,a4paper]{article}

\advance\voffset by -2.0cm \advance\hoffset by -1.25cm
\def\PLA#1#2#3{{ Phys.\ Lett.}\/ {\bf A#1} (#2) #3}
\def\PRD#1#2#3{{ Phys.\ Rev.}\/ {\bf D#1} (#2) #3}

\def\IJMP#1#2#3{{ Int.\ J.\ Mod.\ Phys.}\/ {\bf A#1} (#2) #3}

\usepackage{amsmath}
\usepackage{amssymb}
\usepackage{amsthm}
\usepackage{mathrsfs}
\usepackage{amsfonts}
\usepackage{revsymb}
\usepackage{cancel}

\allowdisplaybreaks[1]
\numberwithin{equation}{section}

\theoremstyle{definition}

\newcommand{\RR}{\mathbb{R}} 
\newcommand{\ZZ}{\mathbb{Z}} 

\DeclareMathOperator{\Tr}{Tr} 

\newcommand{\beq}{\begin{equation}}
\newcommand{\eeq}{\end{equation}}
\newcommand{\beqa}{\begin{eqnarray}}
\newcommand{\beqn}{\begin{eqnarray}}
\newcommand{\eeqn}{\end{eqnarray}}
\newcommand{\eeqa}{\end{eqnarray}}

\def\1{\frak 1}
\def\2{\frak 2}
\def\3{\frak 3}

\newcommand{\HH}{\mathcal{H}}

\newlength{\oldcolsep}

\begin{document}

\title{The Geometrical Origin of Dark Energy}

\author{Alon E. Faraggi$^1$ and Marco Matone$^{2,3}$}\date{}

\maketitle

\begin{center}{\it $^1$Department of Mathematical Sciences\\
University of Liverpool,
Liverpool L69 7ZL, UK }\end{center}

\begin{center}{\it $^2$Dipartimento di Fisica e Astronomia ``G. Galilei'' \\
Universit\`a di Padova, Via Marzolo 8, 35131 Padova,
Italy}\end{center}

\begin{center}{\it $^3$INFN, Sezione di Padova\\  Via Marzolo 8, 35131 Padova,
Italy}

\end{center}

\bigskip

\begin{abstract}
The geometrical formulation of the quantum Hamilton-Jacobi theory
shows that the quantum potential is never trivial, so that it plays the r\^ole of intrinsic energy.
Such a key property selects the Wheeler-DeWitt (WDW) quantum potential $Q[g_{jk}]$ as the natural candidate
for the dark energy. This leads
to the WDW Hamilton-Jacobi equation
with a vanishing kinetic term, and with the identification
$$
\Lambda=-\frac{\kappa^2}{\sqrt{\bar g}}Q[g_{jk}] \ .
$$
This shows that the cosmological constant is a quantum correction of the Einstein tensor, reminiscent
of the von Weizs\"acker correction to the kinetic term of the Thomas-Fermi theory. The quantum potential also defines the Madelung pressure tensor.
The geometrical origin of the vacuum energy density, a strictly non-perturbative phenomenon,
provides strong evidence that it is due to a graviton condensate. Time independence of the regularized WDW equation
suggests that the ratio between the Planck length and the Hubble radius may be a time constant, providing an infrared/ultraviolet duality.
We speculate that such a duality is related to the local to global geometry theorems for constant curvatures, showing that understanding
the universe geometry is crucial for a formulation of Quantum Gravity.

\end{abstract}

\noindent

\newpage
\setcounter{footnote}{0}
\renewcommand{\thefootnote}{\arabic{footnote}}

\section{Introduction}

In spite of the tremendous efforts, understanding the origin of the cosmological constant
\cite{Weinberg:1988cp}\cite{Carroll:2000fy}\cite{Sola:2013gha} is still an open question.
In this paper we show that the cosmological constant is naturally interpreted in terms of the
quantum potential associated to the spatial metric tensor. The starting point concerns the geometrical formulation of the Quantum Hamilton-Jacobi Equation (QHJE), suggested by the $x-\psi$ duality observed in \cite{Faraggi:1996rn}
and introduced in \cite{fm2} (see \cite{Faraggi:2009ra} for a short review).
In the following we call such a formulation, which differs with respect to the Bohmian one, {\it Geometrical Quantum Hamilton-Jacobi} (GQHJ) theory.
Such a theory reproduces the main results of Quantum Mechanics (QM), including energy quantization and tunneling, without
using any probabilistic interpretation of the wave function, which is one of the problems in formulating a consistent theory of quantum gravity.

\noindent
Another consequence of the GQHJ theory is that if space is compact, then there is no notion of particle trajectory \cite{Faraggi:2012fv}.
It follows that the GQHJ theory reproduces the results of QM
following a geometrical approach without the axiomatic interpretation of the wave function as probability amplitude.

\noindent
The idea underlying the geometrical derivation of the QHJE is that, like General Relativity (GR), even QM has a geometrical interpretation. This is done by imposing
the existence of point transformations connecting different states, which, in turn, leads to a cocycle condition that uniquely fixes the
QHJE. It is then immediate to show that the QHJE implies the Schr\"odinger equation.
In such a formulation, it has been shown that the quantum Hamilton characteristic function $S$ is non-trivial even in the case of the free particle
with vanishing energy. Such a result is deeply related to the solution of Einstein's paradox, discussed later, and concerning the classical limit of bound states in the de Broglie-Bohm theory.

\noindent
In the present paper we are interested in the fact that, unlike in the de Broglie-Bohm theory, the quantum potential in the GQHJ theory is never trivial \cite{fm2}.
This happens even in the case of a free particle with vanishing energy.
It is just such a property that led in \cite{Matone:2000ge} to the proposal that there is a deep relation between
QM and gravity. In particular, it was emphasized that the characteristic property of the quantum potential is its universal nature, which is, like gravity, a
property possessed by all forms of matter. Subsequently, the deep relation between gravity and QM was
also stressed by Susskind in his GR=QM paper \cite{Susskind:2017ney} and where it is emphasized that
where there is quantum mechanics there is also gravity. An explicit relation between quantum mechanics and gravity
arises in the case of the free particle with vanishing energy, whose quantum potential includes the Planck length $\ell_P=\sqrt{\hbar G/c^3}$
\cite{Matone:2000ge}
\beq
Q(x)=\frac{\hbar^2}{4m}\{S,x\}=-\frac{\hbar^2}{2m}\frac{\ell_P^2}{(x^2+\ell_P^2)^2}   \ ,
\label{nelcasolibero}\eeq
where $\{f,x\}=f'''/f' - \frac{3}{2}(f''/f')^2$ is the Schwarzian derivative of $f$.
Such a result follows by requiring that, in the case  of a free particle of energy $E$, the QHJE
consistently reproduces both the $\hbar\to0$ and $E\to0$ limits. On the other hand, since in the problem
there are no scales, one is forced to use universal constants. It turns out that the Planck length is the only candidate satisfying the limit conditions,
a result related to the invariance of the quantum potential under M\"obius transformations of $S$.
Since $E=0$ corresponds to the ground state, it follows that $Q$ can be considered as an
intrinsic energy.

\noindent
The GQHJ theory includes another relation between QM and geometry of the universe. Namely, compactness of space would imply
that the energy spectra are quantized \cite{Faraggi:2012fv}. The essential reason is that solutions of the Schr\"odinger
equation should satisfy gluing conditions, so implying a quantized spectra, even in the case of the free particle  \cite{Faraggi:2012fv}. This is also connected
to the problem of definition of time. To see this, note that while in classical mechanics we have the equivalence between
the definition of trajectory given by  $p=\vec \nabla S$ and the one following by
Jacobi theorem, that is
\beq
p=\vec \nabla S \quad \longleftrightarrow \quad t-t_0=\frac{\partial S}{\partial E} \ ,
\eeq
at the quantum level the two definitions do not coincide. As shown in \cite{floyd}\cite{Faraggi:2012fv}, trajectories, if any, should be defined by the Jacobi theorem.
On the other hand, since a compact universe implies a quantized energy spectra, it follows that in this case the derivative of $S$ with respect to
$E$ is ill-defined \cite{Faraggi:2012fv}.\footnote{An alternative to the ill-defined derivative $\partial_E S$ is to consider finite differences in the $E-S$ plane. One may easily check that this leads to a heuristic uncertainty relation between $E$ and $t$.}
We then have
\beq
\text{Compact Universe} \longrightarrow \{E_{n}\} \longrightarrow \frac{\partial S}{\partial E} \,\, \text{is ill-defined}
\longrightarrow \text{no notion of trajectories} \ .
\label{unopunottre}\eeq
This leads to a possible relation between the problem of time in GR and the fact that time is not an observable in QM.
It should be stressed that
in Quantum Field Theory (QFT), even particle's spatial position is represented by parameters, so that, like time, even such a notion does not correspond to an observable.

\noindent It is worth mentioning that the GQHJ theory has been inspired by uniformization theory, with the Schr\"odinger equation playing the analogous r\^ole
of the uniformizing equation. In particular, the ratio of two linearly independent solutions of the Schr\"odinger equation,
plays the analogous r\^ole of the inverse of the uniformizing map. The basic duality,
that is the M\"obius symmetry, which extends to the QHJE in higher dimension \cite{bfm}, is the defining property of the Schwarzian derivative. Such a duality,
that relates small and large scales, and acts like the map between different fundamental domains, is
at the heart of the proof of the energy quantization \cite{fm2}. The above connection between compactness of space, discrete spectra and the analogies with
uniformization theory, suggests that higher dimensional uniformization theory is related to the geometry of the universe. This would imply
that Thurston's geometry \cite{Thurston} is the appropriate framework to describe the Universe. In this context, the 3-torus plays a central r\^ole.

\noindent
Besides (\ref{nelcasolibero}), also (\ref{unopunottre})
provides a relation between small and large scales. In particular, as in the case of a particle in a ring of radius $R$, that gives
$E_n = {n^2\hbar^2}/{(2mR^2)}$, $n\in\ZZ$,
an analogous relation shows that the energy spacing depends on the parameters defining the compact geometry of space.

\noindent
We saw that the GQHJ theory indicates that QM and GR are deeply related.
In particular, in the GQHJ theory, time is not a well-defined observable.
On the other hand, in the quantum gravity equation par excellence, that is the Wheeler–DeWitt (WDW) equation \cite{DeWitt:1967yk}\cite{Wheeler}, there is no time variable at all.

\noindent
The above analysis suggests considering the r\^ole of the WDW quantum potential.
In the case of quantum gravity, the quantum potential represents an intrinsic energy density.
In analogy with the GQHJ theory and, in particular, with (\ref{nelcasolibero}),
the natural interpretation is that the WDW quantum potential in the vacuum is
the one of dark energy, that is
\beq
\Lambda =-\frac{\kappa^2}{\sqrt{\bar g}}Q[g_{jk}]\ ,
\label{QeLambdaDensitaEnergiaIntrinsica}\eeq
where $\bar g=\det g_{jk}$.
We then have that the cosmological constant
is a quantum correction to the Einstein tensor. This is reminiscent
of the von Weizs\"acker correction to the kinetic term of the Thomas-Fermi theory \cite{Weizsacker}.
It is worth mentioning that also the Madelung pressure tensor is defined in terms of the quantum potential.

\noindent
Since (\ref{QeLambdaDensitaEnergiaIntrinsica}) refers to the vacuum, it follows that there are no dynamical degrees of freedom,
so that $S=0$. This means that (\ref{QeLambdaDensitaEnergiaIntrinsica})
coincides with the WDW equation in the vacuum.

\noindent
A consequence of our investigation is that since the metric tensor is the only field involved in (\ref{QeLambdaDensitaEnergiaIntrinsica}),
it follows that dark energy is naturally identified with a graviton condensate.
We note that, in a quite different context, the r\^ole of the (Bohmian) quantum potential in cosmology, suggesting that the vacuum is a graviton
condensate, has been proposed in \cite{Das:2014agf}.

\noindent
We will argue that, as suggested by Feng's volume average regularization \cite{Feng:2018cul},
and by the minisuperspace approximation, a regularized WDW equation
would need, besides the Planck length, the addition of an infrared scale that we identify with the Hubble radius
$R_H=c/H_0=1.36\cdot 10^{26}m$.
Time independence of the regularized WDW equation would then imply
that, like $R_H$, even the Planck length is time-dependent. In particular,
time independence of the WDW wave-functional suggests that
\beq
\mathcal{K}=\frac{\cancel\ell_P}{R_H}=5.96\cdot 10^{-61} \ ,
\label{nuovacostante}\eeq
may be a space-time constant. This would provide an exact infrared/ultraviolet duality.

\noindent
The paper is organized as follows.
In sect. 2 we shortly review the derivation of the WDW Hamilton-Jacobi (HJ) equation.
Sect. 3 illustrates the main points of the GQHJ theory
formulated in \cite{fm2}, focusing on its geometrical origin and on the solution of Einstein's paradox, which in turn is related to the non-triviality
of the QHJE for the free particle with $E=0$.
In sect. 4 we show that, contrary to the de Broglie-Bohm formulation, the quantum potential is non-trivial even in the case of the WDW HJ equation
with $^3R=0$ and vanishing cosmological constant.
In sect. 5 we show that the cosmological constant is naturally interpreted in terms of the WDW quantum potential in the vacuum.
We then derive the wave-functional in the minisuperspace approximation. Sect. 6 is devoted to some speculative suggestion
relating the infrared/ultravilet duality, in the context of the regularized WDW equation,
to the local to global geometry theorems concerning manifolds of constant curvature.
It turns out that the global geometry is strongly constrained in case the local
one has constant curvature.
 This is just the geometrical counterpart of the fact that large scale physics seems constrained by the physics at small scales.
Another manifestation of the connection between QM and GR.
Finally, we argue that time independence of the regularized WDW equation would imply that $\mathcal{K}$ is a space-time constant.

\section{WDW Hamilton-Jacobi equation}

 In the Arnowitt, Deser and Misner (ADM) formulation \cite{Arnowitt:1962hi}, the space-time is foliated into a family of closed space-like hypersurfaces parametrized by time.
One then considers such spatial hypersurfaces at ``constant time'',
as level sets of a time function
\beq
\Sigma_{t_0}=\{x^k|t(x^k)=t_0\} \ .
\eeq
In the following we choose the metric signature $(-,+,+,+)$.
Denote by $g_{ij}={}^4 g_{ij}$ the metric tensor of the three dimensional spatial slices.
Let $N=(-{}^4 g^{00})^{-1/2}$ be the lapse and $N_k={}^4 g_{0k}$ the shift vector field. We then have the standard 3+1 decomposition
\beq
ds^2=(N_kN^k-N^2)c^2dt^2+2N_k c dx^k dt+g_{jk} dx^jdx^k \ .
\eeq
Note that $N$, $N_k$ and $g_{jk}$ depend on $(t_0,x^1,x^2,x^3)$.
As we will see, the lapse function and the shift vector field play the role of four Lagrange multipliers and describe the welding of the $\Sigma_{t}$'s.
The equations of motion for $N$ and $N_k$ are arbitrary, reflecting the freedom in choosing the space-time coordinates \cite{Arnowitt:1962hi}\cite{MBlau}\cite{Calcagni:2017sdq}.

\noindent
Set $\bar g=\det g_{jk}$ and $\kappa^2=8\pi G/c^4$. The Einstein-Hilbert Lagrangian density can be equivalently expressed in the form
\beq
{\mathscr L}=\frac{1}{2\kappa^2}N \sqrt{\bar g}({}^3 R-2\Lambda +K^{jk}K_{jk}-K^2) \ ,
\eeq
where ${}^3 R$ is the intrinsic spatial scalar curvature, $\Lambda$ the cosmological constant, $K$ the trace of the extrinsic curvature
\beq
K_{jk}=\frac{1}{N}\Big(\frac{1}{2}g_{jk,0}-D_{(j}N_{k)}\Big)\ ,
\eeq
and $D_j$ denotes the $j$ component of the covariant derivative. Let $\pi^0$ and $\pi^k$ be the
momenta conjugate to $N$ and $N_k$ respectively. Since ${\mathscr L}$ is independent of both
$\partial_{x_0}N$ and $\partial_{x_0}N_k$,
we have the primary constraints
$\pi^0\approx 0$, $\pi^k \approx 0$. Here the symbol ``$\approx$'' indicates weak equality, that is
the vanishing holding only on the sub-manifold of the phase space constrained by the primary constraints. The equality
holding only when the expression is identically vanishing on the full phase space \cite{Calcagni:2017sdq}.

\noindent
Time conservation of the primary constraints implies
secondary constraints, given by the weak vanishing of the super-momentum,
\beq
\HH_k=-2D_j\pi^j_{\;\,k} \approx 0 \ ,
\eeq
and of the super-Hamiltonian,
\beq
\HH=2\kappa^2 G_{ijkl}\pi^{ij}\pi^{kl}-\frac{1}{2\kappa^2}\sqrt{\bar g}({}^3R-2\Lambda)\approx 0 \ ,
\label{dropilprimo}\eeq
where $\pi^{jk}$ is the momentum canonically conjugated to $g_{jk}$, that is
\beq
\pi^{jk}=-\frac{1}{2\kappa^2}\sqrt{\bar g}(K^{jk}-g^{jk}K) \ ,
\label{perdimensioni}\eeq
and
\beq
G_{ijkl}=\frac{1}{2\sqrt{\bar g}}(g_{ik}g_{jl}+g_{il}g_{jk}-g_{ij} g_{kl}) \ ,
\eeq
is the DeWitt supermetric. The conservation in time of the secondary constraints do not imply further constraints.

\noindent
By a Legendre transform one gets the Hamiltonian
\beq
H=\int d^3{\bf x} (N \HH +N^k\HH_k) \ ,
\eeq
showing that $N$ and $N^k$ are the Lagrange multipliers of $\HH$ and $\HH_k$ respectively.

\noindent
Let $\Psi$ be the Schr\"odinger wave-functional, that is
$i\hbar\partial_t\Psi=\hat H \Psi$. Implementation of the primary constraints at the quantum level is obtained by setting
\beq
\hat\pi^0=-i\hbar\frac{\delta}{\delta N} \ , \qquad \hat \pi^k=-i\hbar\frac{\delta}{\delta N_k} \ ,
\eeq
so that
\beq
-i\hbar\frac{\delta\Psi}{\delta N}=0 \ , \qquad -i\hbar\frac{\delta\Psi}{\delta N_k}=0 \ ,
\eeq
meaning that $\Psi$ does not depend on any of the non-dynamical variables, that is $\Psi$ depends
on $g_{jk}$ only.

\noindent
At the quantum level the conjugate momenta of a field $\phi$ would correspond to
$-i\hbar\delta_\phi$, so that,
since for the $\delta$-distribution in configuration space we have $[\delta^{(3)}]=L^{-3}$, it follows that
$[\delta_\phi]=[\phi]^{-1}L^{-3}$.
On the other hand, by (\ref{perdimensioni}) we have $[\pi_{ij}]=MT^{-2}$,
which is different from the dimension of the canonical choice of $\hat\pi^{jk}$, namely
$[-i\hbar\delta_{g_{jk}}]=ML^{-1}T^{-1}$. We then have
\beq
\hat\pi^{jk}=-i\hbar c \frac{\delta}{\delta g_{jk}} \ ,
\label{laquantistica}\eeq
which also fixes the normalization of the classical relation
\beq
\pi^{jk}=c\frac{\delta S}{\delta g_{jk}} \ ,
\label{Sindotta}\eeq
where $S$ is the functional analogue of Hamilton's characteristic function.
By (\ref{laquantistica}), the super-momentum constraint reads
\beq
\hat\HH_k \Psi=2i\hbar c g_{kj}D_l\frac{\delta\Psi}{\delta g_{lj}}=0 \ ,
\eeq
which is satisfied if $\Psi$ is invariant under diffeomorphisms of the hypersurface.

\noindent
The other secondary constraint, that is
$\hat \HH\Psi=0$,
is the WDW equation
\beq
\hbar c\Big[-2\cancel{\ell}_P^2  G_{ijkl}\frac{\delta^2}{\delta g_{ij}\delta g_{kl}}-\frac{1}{2\cancel{\ell}_P^2}
\sqrt{\bar g}({}^3R-2\Lambda)\Big]\Psi[g_{ij}]=0 \ ,
\label{WDWI}\eeq
where $\cancel{\ell}_P=\sqrt{8\pi \hbar G/c^3}=\kappa \sqrt{\hbar c}$ is the rationalized Planck length.
Note that the secondary constraints imply $\hat H\Psi=0$, so that $\partial_t\Psi=0$, which is
the origin of the problem of time.

\noindent
Let us now consider the key identity
\beq
\frac{1}{A{\rm e}^{\beta S}}
\frac{ \delta^2 \left(A{\rm e}^{\beta S}\right)}
{\delta g_{ij} \delta g_{kl}}
=\beta^2
\frac{\delta S}{\delta g_{ij}}
\frac{\delta S}{\delta g_{kl}}
+
\frac{1}{A}
\frac{  \delta^2 A}
{\delta g_{ij} \delta g_{kl}}
+
\frac{\beta}{2A^2}\Bigg[
\frac{\delta}{\delta g_{ij}}
\left(A^2\frac{\delta S}{\delta g_{kl}}\right)+\frac{\delta}{\delta g_{kl}}
\left(A^2 \frac{\delta S}{\delta g_{ij}}\right)\Bigg] \ ,
\label{wdwid}
\eeq
which holds for any complex constant $\beta$.
Set $\beta=i/\hbar$ and
\beq
\Psi=A e^{\frac{i}{\hbar}S} \ ,
\label{PsinoBohm}\eeq
with $A$ and $S$ taking real values. In this respect, note that if $A e^{\frac{i}{\hbar}S}$ is a solution, then reality of the WDW operator
implies that even $A e^{-\frac{i}{\hbar}S}$. This observation is related to the differences between the Bohmian and the GQHJ formulations discussed later.
Replacing $\Psi$ in (\ref{WDWI}) with right hand side of (\ref{PsinoBohm}) gives the WDW HJ equation, corresponding to the following quantum deformation of the
HJ equation
\beq
2(c\kappa)^2 G_{ijkl}
\frac{\delta S}{\delta g_{ij}}
\frac{\delta S}{\delta g_{kl}}-
\frac{1}{2\kappa^2}\sqrt{\bar g}({}^3 R-2\Lambda)-2(c\kappa\hbar)^2
\frac{1}{A}
G_{ijkl}
\frac{  \delta^2 A}
{\delta g_{ij} \delta g_{kl}}
=0 \ ,
\label{wdwhje}\eeq
together with the continuity equation
\beq
G_{ijkl}\frac{\delta}{\delta g_{ij}}\Big( A^2\frac{\delta S}{\delta g_{kl}}\Big) = 0 \ .
\eeq
The last term in (\ref{wdwhje}), that is
\beq
Q= -2(c\kappa\hbar)^2
\frac{1}{A}
G_{ijkl}
\frac{  \delta^2 A}
{\delta g_{ij} \delta g_{kl}}\ ,
\label{QPDef}\eeq
is called quantum potential. We note that in the classical limit Eq.(\ref{wdwhje}) reduces to the classical case (\ref{dropilprimo}).

\noindent
A key difference between the GQHJ formulation and the Bohmian one
is that, as in the formulation of QM, the $\Psi$ in (\ref{PsinoBohm}) is not in general identified with the
wave-functional of the state of the system, rather it is a general solution of the WDW equation.
In the next section, we will see that it is precisely such a characteristic of the GQHJ formulation that, unlike the Bohmian one,
implies that

\begin{enumerate}

\item there is no Einstein's paradox,

\item there is a basic M\"obius symmetry, associated to the Schwarzian equation,

\item energy quantization follows without the need of any interpretation of the wave-function,

\item implies that in compact space there is no notion of particle trajectory.

\end{enumerate}

\noindent An explicit example of the difference between the GQHJ and Bohmian formulations associated to the WDW HJ equation is provided in
sect. \ref{WDWHJsec}. In particular,
we will consider the case
${}^3R=0$, $\Lambda=0$,
so that the WDW equation reduces to the free functional differential equation. While in the Bohmian formulation
this would imply
\beq
\Psi=0 \ ,
\eeq
so giving $A=0$, $S=0$ and $Q=0$, in the GQHJ theory there are non-trivial solutions. Once again, this shows that,
contrary to the Bohmian formulation, the quantum potential in the GQHJ theory is never trivial, so that
it plays the r\^ole of intrinsic energy.

\section{QHJE and Einstein paradox}

In this section we shortly discuss the main aspects of the GQHJ theory \cite{fm2}.
Let us start by recalling Einstein's paradox
(see e.g. Ref. \cite{Hollandbook} pg. 243). This concerns the issue in Bohmian mechanics when considering the classical limit
for states described by a wave-function corresponding to Hamiltonian eigenstates of any one-dimensional bound state.
Let us then consider a state of definite energy $E$ and denote by $\psi_E$ the corresponding wave-function. In this case one can easily
show that $\psi_E\in L^2(\RR)$ is proportional to a real function. Therefore, if one sets, as in Bohm theory, $\psi_E=R e^{\frac{i}{\hbar}S}$, then $S$ is a constant. On the other hand, in the
Bohmian formulation,
$p=\partial_x S$ is identified with the mechanical
momentum $m\dot x$, so that, quantum mechanically, one would have $p=0$. This would imply that, as in the case of the harmonic oscillator, a quantum particle would be at rest and
should start moving in the classical limit, where $S$
and $p$ are non-trivial. In other words, it is clear that it is not possible to get a non-trivial $S$ as the $\hbar\to0$ limit of $S=0$.

\noindent
The resolution of the paradox is that the quantum analogue of $S$
is not necessarily the phase of the wave function. As we will show, this in fact also
underlies the WKB approximation that, even if one starts with the identification $\psi=\exp(iS_{WKB}/\hbar)$, with $S_{WKB}$ complex, then
real wave functions are identified with a linear combination of in and out waves.
In our formulation, such a choice is not ad hoc as in the WKB approximation, rather it follows from the request that
the cocycle condition is always satisfied \cite{fm2}. In particular, note that
if $R e^{\frac{i}{\hbar}S}$ is a solution of the stationary Schr\"odinger equation (SSE), then, this is also the case of $R e^{-\frac{i}{\hbar}S}$. This is the key to introduce the so-called bipolar decomposition
\beq
\psi_E=R\Big(Ae^{\frac{i}{\hbar}S}+Be^{-\frac{i}{\hbar}S}\Big) \ ,
\label{Einsteinparadox}\eeq
which is equivalent to say
that the most general expression for $S$, and therefore
for $R$, is given by
\beq
Re^{\frac{i}{\hbar}S}=A\psi^D+B\psi \ ,
\label{equivv}\eeq
with $\psi^D$ and $\psi$ two arbitrary linearly independent solutions of the SSE.

\noindent
As a result, in the case of a real $\psi_E$, the only constraint is just $|A|=|B|$ and one gets a non-trivial $S$ with a well-defined classical limit. Such a solution
of Einstein's paradox is a consequence of the GQHJ theory, that excludes in a natural way, and from the very beginning, the existence of states
with a constant $S$ \cite{fm2}. The use of the bipolar decomposition was previously discussed by Floyd \cite{floyd}.

\noindent
Later we will see that in the case of the WDW HJ equation, both $S$ and the quantum potential
are non-trivial even when ${}^3 R=0$ and $\Lambda=0$.
This is the functional analogue of basic properties of the quantum potential in the GQHJ theory
that we now discuss.

\noindent
The main point that characterizes the non-trivial properties of the quantum potential is its connection
with the M\"obius invariance of the Schwarzian derivative $\{f,x\}$,
that, in order to be well-defined, requires that $f\in C^2(\RR)$ and $\partial_x^2 f$ differentiable on $\RR$.
The continuity equation $\partial_x(R^2\partial_x S)=0$ implies that $R$ is proportional to $(\partial_x S)^{-1/2}$, so that
the quantum potential can be expressed in terms of $S$ only
\beq
Q=\frac{\hbar^2}{4m}\{S,x\} \ ,
\eeq
and the QHJE associated to a SSE reduces to the single equation
\beq
\frac{1}{2m}\Big(\frac{\partial S}{\partial x}\Big)^2+ V-E+Q=0 \ .
\label{okd}\eeq
 Let us consider the basic identity
\beq
\left(\frac{\partial S}{\partial x}\right)^2=
\frac{\beta^2}{2}
\left(\left\{{\rm e}^{\frac{2i}{\beta}{S}},x\right\}-
\left\{S,x\right\}\right) \ ,
\label{schwarzianidentity}
\eeq
where $\beta$ is a constant with the dimension of an action.
Such an identity implies that the QHJE (\ref{qhjee}) can be also expressed in the form
\beq
\left\{\exp\Big(\frac{2i}{\hbar}S\Big),x\right\}=\frac{4m^2}{\hbar}(E-V) \ .
\label{qhjee}\eeq
The solution of this non-linear differential equation is
\beq
\exp\Big(\frac{2i}{\hbar}{S}\Big)=\gamma\Big[\frac{\psi^D}{\psi}\Big] \ ,
\eeq
where
$\psi$ and $\psi^D$ are two real linearly independent solutions of the SSE
and $\gamma[f]$ is an arbitrary, generally complex, M\"obius transformation of $f$
\beq
\gamma [f]= \frac{Af+B}{Cf+D} \ .
\eeq
Thanks to the M\"obius invariance of the Schwarzian derivative, one may consider
a M\"obius transformation of $\exp(2iS/\hbar)$, that we denote again by
\beq
\gamma\Big[{\exp}\Big(\frac{2i}{\hbar}{S}\Big)\Big] \ ,
\eeq
leaving $V-E$ invariant. On the other hand, since this corresponds to the transformation
\beq
S \longrightarrow \tilde S = \frac{\hbar}{2i}\log \gamma\Big[{\exp}\Big(\frac{2i}{\hbar}{S}\Big)\Big] \ ,
\eeq
we see that there is a non-trivial mixing between the kinetic term and the quantum potential in (\ref{okd}).

\noindent
In \cite{fm2} the QHJE was derived by a slight modification of the way one gets the classical HJ equation.
Namely, instead of looking for maps from $(x,p)$ to $(X,P)$, seen as independent variables, such that the new Hamiltonian is the trivial one, $\tilde H=0$, we looked for transformations
$x\to \tilde x$ such that $\tilde V-\tilde E=0$, but with the transformation of $p$ fixed by imposing that $S(x)$ transforms as a scalar function. We then have
\beq
\tilde S(\tilde x)=S(x) \ ,
\label{lacondizione}\eeq
holding for any pair of physical systems, including the one with $V-E=0$.

\noindent
A key consequence of (\ref{lacondizione}) is that  $S(x)$ can never be a constant. In particular, imposing that (\ref{lacondizione}) holds even when the coordinate $x$ refers to the state with $V-E=0$,
forces the introduction of an additional term
in the classical HJ equation. Then, one considers three arbitrary states, denoted by $A$, $B$ and $C$,
and imposes the condition coming from the commutative diagram of maps
$$
\begin{array}{c} {} \\ {} \\ A\end{array}
\begin{array}{c} {} \\ \nearrow \\ {} \end{array}
\begin{array}{c} B \\ {} \\ \longrightarrow \end{array}
\begin{array}{c} {} \\ \searrow \\ {} \end{array}
\begin{array}{c} {} \\ {} \\ C\end{array}
$$
Implementation of such a consistency condition is equivalent to a cocycle condition that fixes the additional term to be the quantum potential
\cite{fm2}. The outcome is just the QHJE.

\noindent
Another feature of the above formulation is that the quantum potential is never trivial even in the case $V-E=0$. In particular,
a careful analysis of the quantum potential for a free particle with vanishing energy shows that the $\hbar\to0$ and $E\to0$
limits in the case of the free particle of energy $E$, leads to the appearance of the Planck length in the expression for the quantum potential $Q$ of a free particle with $E=0$, given in Eq.(\ref{nelcasolibero}).
It should be stressed that the present formulation leads to a well-defined power expansion in $\hbar$ for $S$. This is different with respect to the WKB approximation since $S_{\rm WKB}$ is defined by
\beq
\psi=\exp\Big(\frac{i}{\hbar}S_{\rm WKB}\Big) \ ,
\eeq
so that, in general, $S_{\rm WKB}$ takes complex values. The GQHJ theory is also different with respect to the de Broglie-Bohm theory.
Besides the case of real wave-functions illustrated above, also the quantum potential (\ref{nelcasolibero}) turns out to be different. The difference also appears in the case of the free particle of energy $E$.
Indeed, the solution of Eq.(\ref{okd}) with $V=0$ is
\beq
S = \frac{\hbar}{2i}\log\left(\frac{Ae^{\frac{2i}{\hbar}\sqrt{2mE}x}+B}
{Ce^{\frac{2i}{\hbar}\sqrt{2mE}x}+D}\right) \ .
\label{indice}\eeq
Here the constants are chosen in such a way that $S\neq\pm\sqrt{2mE}x$. Such a choice, fixed by the consistency condition
that the non-trivial $S_{E=0}$ is obtained from $S$ in the $E\to0$ limit, relates $p$-$x$ duality, also called Legendre duality, and M\"obius invariance of the Schwarzian derivative \cite{fm2}.
Another consistency condition comes from the classical limit. Since $S^{cl}=\pm\sqrt{2mE}x$, we have
\beq
\lim_{\hbar\longrightarrow0}\log\left(\frac{Ae^{\frac{2i}{\hbar}\sqrt{2mE}x}+B}
{Ce^{\frac{2i}{\hbar}\sqrt{2mE}x}+D}\right)^{\frac{\hbar}{2i}}={\pm\sqrt{2mE}x} \ ,
\label{alluce}\eeq
implying that the constants $A$, $B$, $C$ and $D$  depend on $\hbar$ \cite{fm2}.

\noindent
The above analysis shows that $S$ is the natural quantum analog of the classical
Hamiltonian characteristic function. The formulation solves Einstein's paradox and the power expansion of $S$ in $\hbar$ is completely under control.
Furthermore, it leads to a dependence of $S$
on the fundamental constants, shedding light on the quantum origin of interactions. It also implies that if space is compact, then
time parametrization cannot be defined \cite{Faraggi:2012fv}.
The formulation, that follows from the
simple geometrical principle (\ref{lacondizione}), extends to arbitrary dimensions and to the relativistic case as well \cite{bfm}. It
reproduces, together with other features, such as energy quantization, the non existence of trajectories,
without assuming any interpretation of the wave-function.

\section{The WDW HJ equation with ${}^3R=0$ and $\Lambda=0$}\label{WDWHJsec}

Let us go back to the WDW equation by considering the case
${}^3R=0$, $\Lambda=0$
\beq
G_{ijkl}\frac{\delta^2}{\delta g_{ij}\delta g_{kl}}\Psi=0 \ .
\label{WDW2AS}\eeq
Setting $\Psi=Ae^{\frac{i}{\hbar}S}$, the WDW HJ equation reads
\beq
G_{ijkl}
\frac{\delta S}{\delta g_{ij}}
\frac{\delta S}{\delta g_{kl}}-
\frac{\hbar^2}{A}
G_{ijkl}
\frac{  \delta^2 A}
{\delta g_{ij} \delta g_{kl}}=0 \ .
\label{wdwhjewithzero}
\eeq
As shown in the previous section, a key difference between the GQHJ formulation and the Bohmian one, is that
in the latter $Re^{\frac{i}{\hbar}S}$ is identified with the wave-function describing the physical state. This is not in general
the case in the GCHJ formulation.
In the Bohmian interpretation, the only admissible solution of Eq.(\ref{WDW2AS}) is the one where the wave-functional is trivial,
so that, as in the case of the free particle with $E=0$, one would have $\Psi=0$, implying $A=0$, $S=0$ and $Q=0$. In the following we show
that, as in the case of (\ref{nelcasolibero}), the general solution of (\ref{WDW2AS}) implies non-trivial $A$, $S$ and $Q$.

\noindent
Note that in the case of (\ref{WDW2AS}) the formulation does not suffer the well-known problem of the WDW equation, due to the
presence of the second-order functional derivative at the same point: such an operator is in general ill-defined since it may lead to
$\delta^{(3)}(0)$-singularities.
On the other hand, the wave functional $\Psi[g_{ij}]$ now depends linearly on
$g_{ij}$, so that the action of the second-order functional derivative
on $\Psi[g_{ij}]$ is well-defined. We then have
\beq
\Psi[g_{ij}]=A{\rm e}^{\frac{i}{\hbar} S}={\cal T}  g + C \ ,
\label{anchequestasoluzione}\eeq
where
\beq
{\cal T}  g:=\int d^3{\bf x} {\cal T}^{jk}({\bf x})g_{jk}({\bf x}) \ ,
\eeq
with ${\cal T}_{jk}({\bf x})$ an arbitrary complex tensor density field of weight 1 and $C$ a complex constant.
\noindent
The most general expression of $S$ is
\beq
\exp\Big(\frac{2i}{\hbar}S\Big)=\frac{{\cal T}  g + C}{\bar{\cal T} g + \bar C} \ ,
\label{esponenzialeazione}\eeq
and for $A$ we have
\beq
A=|{\cal T}  g + C| \ .
\eeq
By (\ref{Sindotta}) and (\ref{esponenzialeazione}), it follows that at the quantum level the momentum conjugate to $g_{jk}$ is
\beq
\pi^{jk}=c\frac{\delta S}{\delta g_{jk}({\bf x})}=\hbar c\,{\rm Im}\,\Big(\frac{{\cal T}^{jk}({\bf x})}{{\cal T} g + C} \Big) \ ,
\label{momento}\eeq
so that the kinetic term in the WDW HJ equation reads
\begin{align}
2(c\kappa)^2G_{ijkl} & ({\bf x})\frac{\delta S}{\delta g_{ij}({\bf x})}\frac{\delta S}{\delta g_{kl}({\bf x})} \cr
& =\frac{2(c\kappa\hbar)^2}{\sqrt {\bar g}}
\Big(\frac{{\cal T}_{kl}({\bf x})}{{\cal T}g+C}\Big) {\rm Im}\,\Big(\frac{{\cal T}^{kl}({\bf x})}{{\cal T}g+C}\Big)
-\frac{1}{2}\Big[{\rm Im}\,\Big(\frac{\Tr{\cal T}({\bf x})}{{\cal T}g+C}\Big)\Big]^2\Big\} \ .
\label{ilcinetico}\end{align}
Note that, by (\ref{wdwhjewithzero}), this also corresponds to $- Q[g_{jk}]$. Furthermore, one may easily check that
such an expression of $Q[g_{jk}]$
is just the functional analogue of the quantum potential of the free particle of vanishing energy
(\ref{nelcasolibero}).

\section{Cosmological constant from the quantum potential}

The discrepancy between the measured value of the cosmological constant and the theoretical prediction
follows by considering $\Lambda/\kappa^2$ as a contribution to the effective vacuum energy density $\rho_{eff}=\rho+\Lambda/\kappa^2$,
where $\langle T_{\mu\nu}\rangle=\rho g_{\mu\nu}$. Considering the QFT vacuum energy density as due to infinitely many
zero-point energy of harmonic oscillators, we get (here $\hbar=c=1$)
\begin{equation}
\rho=\int_{0}^{\Lambda_{UV}}\frac{4\pi k^2 dk}{(2\pi)^3}\frac{1}{2}\sqrt{k^2+m^2}\approx \frac{\Lambda_{UV}^4}{16\pi^2}\approx 10^{71} {\rm GeV}^4 \ ,
\label{rho1}\end{equation}
where  $\Lambda_{UV}$ is the Planck mass. A result which is in complete disagreement with the estimation,
 based on experimental data, $\rho_{eff}\approx 10^{-47} {\rm GeV}^4$.

\noindent
A problem with the above derivation is that it is based on the perturbative formulation of QFT. This corresponds to
use the canonical commutation relations of the free theory that selects the vacuum of the free theory. On the other hand, the true vacuum of
non-trivial QFT's is highly non-perturbative and is not unitarily equivalent to the free one.
As a matter of fact, perturbation theory erroneously treats the quantum fields
evolving as the free ones between point-like interaction events. From the physical point
of view, the r\^ole of renormalization is to iteratively change the parameters of the theory,
that then will depend on the physical scale. In other words, perturbation theory is a
way to mimic the interacting theory by a free one, with the parameters becoming scale dependent.

\noindent
It has been observed in \cite{Cohen:1998zx} that the cutoff corresponding to the value of the cosmological constant
may be related to an infrared/ultraviolet duality. In particular, the authors of \cite{Cohen:1998zx}, inspired by
the Bekenstein bound $S\lesssim  \pi M_P^2L^2$ for the total entropy in a volume of size $L^3$,  proposed the following relation between the infrared cutoff $1/L$ and $\Lambda_{UV}$
\beq
L^3 \Lambda_{UV}^4\lesssim L M_P^2 \ .
\eeq
An estimation of the infrared scale of QFT can be derived by considering the precision tests of the electron's anomalous magnetic moment $a_e$.
In this respect, as observed in \cite{Carmona:2000gd}, an estimate of the correction to the usual calculation
imposed by the IR scale $\mu$ is
\beq
\delta a_e \approx \frac{\alpha}{\pi}\Big(\frac{\mu}{m_e}\Big)\approx 4\cdot 10^{-9} \frac{\mu}{1 {\rm eV}}\ .
\eeq
Requiring that such an indeterminacy be smaller than the uncertainty of the theoretical prediction for $a_e$ gives
\beq
\mu\leq 10^{-2}\, {\rm eV} \ ,
\eeq
which is the value corresponding to the cutoff that
leads to the same order of magnitude of the experimental value of $\rho$.

\noindent
The above analysis indicates that the cosmological constant is related to the infrared problem,
a non-perturbative phenomenon concerning the structure of the vacuum which has physically measured
consequences. For example, QED finite transition amplitudes are obtained by summing over states with infinitely many soft photons.

\noindent
We saw that, unlike in Bohmian mechanics, the quantum potential is never trivial \cite{fm2}.
 This is the case even for the free particle of vanishing energy, implying that the quantum potential plays the r\^ole of
particle intrinsic energy. Furthermore, Eq.(\ref{nelcasolibero}) shows that the quantum potential includes the Planck length, which arises
by consistency conditions in considering the $E\to0$ and $\hbar\to0$ limits \cite{Matone:2000ge}. This was one of the reasons
suggesting a strict relationship between QM and GR \cite{Matone:2000ge} (see also \cite{Susskind:2017ney}).
We then have the following result:

\bigskip

\noindent
{\it The WDW quantum potential in the vacuum corresponds to an intrinsic energy density.}

\bigskip

\noindent It is then natural to make the identification
\beq
Q[g_{jk}]=-\sqrt{\bar g}\rho_{\rm vac} \ ,
\label{laformulafinale}\eeq
$\rho_{\rm vac}=\Lambda/\kappa^2$.
Since in this case the only degrees of freedom are the ones associated to the metric tensor, the dark
energy should correspond to a graviton condensate.

\noindent
In this context, we stress that the
vacuum energy is a purely quantum property and the absence of the kinetic term does not imply, as in the de Broglie-Bohm theory, Einstein's paradox.
The fact that the cosmological constant is a quantum correction to the Einstein tensor given in terms
of the quantum potential
is reminiscent of the von Weizs\"acker correction to the kinetic term of the Thomas-Fermi theory.
Furthermore, we note that the quantum potential also defines the Madelung pressure tensor.

\noindent
Now observe that the absence of propagating degrees of freedom implies that the quantum potential in (\ref{laformulafinale})
corresponds to the one of the WDW HJ equation without the kinetic term, that is
\beq
S=0 \ .
\label{esse}\eeq
Let us choose a metric with vanishing $^3 R$. Eq.(\ref{esse}) implies a nice mechanism, namely
by (\ref{wdwhje}) it follows that in this case the continuity equation is trivially satisfied, so that
Eq.(\ref{laformulafinale}), that by (\ref{esse}) is the full WDW HJ equation, coincides
with the WDW equation (\ref{WDWI}) with $\Psi=A$. In this way the contribution to the WDW HJ equation comes only from the quantum potential.
In other words, since by (\ref{esse}) $\Psi$ takes real values, it follows by the definition of $Q[g_{ij}]$ in (\ref{QPDef}), that Eq.(\ref{laformulafinale}) is just the
WDW equation in the vacuum
\beq
-2\cancel{\ell}_P^2  G_{ijkl}\frac{\delta^2}{\delta g_{ij}\delta g_{kl}}A= -\frac{\sqrt{\bar g}}{\cancel{\ell}_P^2}\Lambda A \ .
\label{WDWI222}\eeq
Note that such an equation is just the functional analog of a stationary Schr\"odinger equation with negative energy.
This suggests considering the role of fundamental scales.
To this end we adapt the analysis that led to Eq.(\ref{nelcasolibero}), to the case of Eq.(\ref{WDWI222}). The main difference is that now the problem
includes both small and large scales.
To see how fundamental constants may appear in the present context, we
first derive an explicit solution of Eq.(\ref{WDWI222}) in the case of the Friedmann-Lema\^{\i}tre-Robertson-Walker background.

\noindent
Let us then consider the line element
\beq
ds^2=-N(t)^2c^2dt^2+a^2(t) d\Sigma_k^2 \ ,
\eeq
where
\beq
d\Sigma_k^2=\frac{dr^2}{1-kr^2}+r^2(d\theta^2+\sin^2\theta d\phi^2) \ ,
\eeq
is the spatial line element of constant curvature $k$. In such an approximation the Hilbert-Einstein equation in the vacuum,
with $k=0$,
reads
\beq
S_{HE}=\frac{V_0}{\kappa^2}\int dt\Big(-\frac{3 a {\dot a}^2}{Nc}-{N c a^3\Lambda}\Big) \ ,
\eeq
where
\beq
V_0=\int dr d\theta d\phi r^2\sin\theta \ .
\eeq
In such a minisuperspace approximation, the WDW equation (\ref{WDWI222}) reads
\beq
\Bigg(\frac{d^2}{da^2}+12 \frac{V_0^2\Lambda}{\cancel{\ell}_P^4}a^4\Bigg)A_{FLRW}=0 \ ,
\label{veramentemini}\eeq
whose solution is a linear combination of the Bessel functions of first and second kind
\beq
A_{FLRW}(a)=\sqrt a\Big(\alpha J_{1/6}(Ca^3)+\beta Y_{1/6}(Ca^3)\Big) \ ,
\eeq
where
\beq
C=2\frac{V_0}{\cancel{\ell}_P^2}\sqrt{\frac{\Lambda}{3}} \ .
\label{lambdaandc}\eeq

\noindent
In this approximation of the WDW equation, besides the Planck length, there is also another fundamental constant, $\Lambda$ itself, and a natural choice, suggested
by (\ref{lambdaandc}), would be
\beq
V_0=\Lambda^{-3/2} \ .
\eeq
Such a result provides an indication on the possible appearance of scales related to the WDW equation. Nevertheless, the analysis should be done in the framework
of the original WDW equation, not just considering its minisuperspace approximation. A key aspect is that the WDWW equation is ill-defined, in particular
it must be regularized, a problem which is completely missing in the minisuperspace approximation.
In the following, we will see that a fundamental scale may in fact appear as an infrared regulator.
In agreement with Dirac's idea, we then will suggest that fundamental constants may be dynamical variables.

\section{Infrared/ultraviolet duality and local to global geometry theorems}

In this section we make some speculation concerning the infrared/ultraviolet duality in the context of the WDW equation, which is
the natural framework to investigate the relations between the structure of the Universe and small scales.
We saw that such an equation includes both large and small scales that can be interpreted as infrared and ultraviolet
cutoffs, that should appear in a well-defined version of the WDW equation. It is clear that such an investigation should include a careful
analysis of the involved local and global geometries.

\noindent
A well-known problem with the WDW equation, is that
due to the second-order functional derivative evaluated at the same point, it presents, in general, $\delta^{(3)}({\bf x}={\bf 0})$-singularities.
 This is analogous to the normal ordering singularities in QFT, due to the joining of two legs of the same vertex; so giving the Feynman propagator evaluated at 0.
Similarly, the infinite volume limit can be interpreted as
the integral representation of the $\delta$-distribution in momentum space at zero momentum. In other words,
$\delta^{(3)}({\bf p}={\bf 0})$
can be interpreted as the infinite space volume limit divided
by $(2\pi)^3$. A related method is used, for example,
in deriving the effective action for $\lambda\phi_4^4$ in Euclidean space  to get the dependence of the coupling constant on the mass scale.
In that case, the infrared regularization was done by supposing that the Euclidean space is $S^4$
rather than $\RR^4$, and then considering $S^4$
as the surface of a five-dimensional sphere, so that one obtains a finite result and avoids such an infrared divergence.

\noindent The outcome of such an analysis is that in general singularities may be removed by taking into account the physical scales. What is crucial is to preserve diffeomorphism invariance. In this respect, we recall
that $[-i\hbar\delta_{g_{jk}}]=ML^{-1}T^{-1}$, while for the $\delta$-distribution in configuration space we have
$[\delta^{(3)}]=L^{-3}$. This means that, besides the Planck length, a well-defined regularized version of the WDW equation
should also involve a large scale cutoff.
An explicit example of such a mechanism
is the one in the interesting paper by Feng, who proposed the volume average regularization \cite{Feng:2018cul}.
Feng's regularization introduces a
factor $1/V$, with $V$ naturally identified with the space volume. In particular, Feng's regularized WDW equation
has the structure
\beq
\hat \HH[\cancel{\ell}_P,V,\Lambda;g_{ij}]\Psi[g_{ij}]=0 \ ,
\eeq
see, for example, Eq.(2.24) of \cite{Feng:2018cul}.
Feng's regularization is related to the standard heat kernel and point splitting regularizations \cite{Wald:1978pj}-\cite{KowalskiGlikman:1996ad}.
In particular, it corresponds to averaging the displacement in the point splitting regularization.

\noindent
The dual r\^ole of the $\delta$-distributions in $x$ and $p$ spaces, shows that infrared and ultraviolet dualities are related to $x$-$p$ duality,
another manifestation of the dual property of the Fourier transform, which in fact is at the heart of the Heisenberg uncertainty relations.

\noindent Even if a well-defined version of the WDW functional differential equation is still unknown, it is clear that,
as Eq.(\ref{veramentemini}) shows, besides the Planck length it should also include a cosmological scale, making manifest
an infrared/ultraviolet duality. A related issue concerns the M\"obius symmetry of the Schwarzian derivative.
In this respect, it was shown in \cite{bfm} that even in the geometrical
 derivation of the QHJE in higher dimensions, there is an
underlying global conformal symmetry, the generalization of the M\"obius symmetry of the Schwarzian derivative.
This is a crucial property,
whose implementation requires a compact space, which in turn would imply that the energy spectra are quantized \cite{Faraggi:2012fv}.
As a consequence, since by Jacobi theorem \cite{floyd}
\beq
t-t_0=\frac{\partial S}{\partial E} \ ,
\eeq
it follows that time-parametrization is ill-defined for discrete spectra, so that trajectories would never exist if space is compact
\cite{Faraggi:2012fv}. The mentioned conformal transformation includes the
space inversion relating large and small scales
\beq
x_k\to l^2 x_k/r^2 \ ,
\eeq
$r^2=\sum_1^D x_k^2$, with $l$ a length scale.
This is another hint that an infrared/ultraviolet duality should appear in the cosmological context, and then
in a well-defined version of the WDW equation. A similar situation arises in the uniformization theory by Klein, Koebe and Poincar\'e,
where negatively curved Riemann surfaces have fundamental domains in their universal covering, e.g. the upper half-plane $\mathbb{H}$,  which are related by
Fuchsian transformations, that is discrete subgroups of\footnote{This is in fact deeply related to the weak/strong duality
transformations of the effective coupling constant $\tau\to -1/\tau$ of Seiberg-Witten theory, that, in the case of pure ${\rm SU}(2)$, posses a $\Gamma(2)\subset{\rm SL}(2,\RR)$ symmetry.} ${\rm SL}(2,\RR)$.

\noindent
Finding an infrared/ultraviolet duality in the cosmological context could be used to consider the
local to global theorems relating local and global geometries. In particular, according to Thurston \cite{Thurston},  the global geometry
is strongly constrained in case the local one has constant curvature. Interestingly, according to Bieberbach \cite{BIeberbach1}\cite{BIeberbach2}, all
compact flat manifolds are finitely covered by tori, a result that in three dimension was previously obtained by Schoenflies \cite{Schoenflies}.
The underlying idea is that the local structure of space provides information on its global structure, which includes the information on the topological structure
and on points at large distances.

\noindent
The discussed connection between compactness of space, discrete
spectra and the analogies with uniformization theory, suggests that higher dimensional
uniformization theory is the right framework to investigate the geometry of the universe.

\noindent
It is clear that the solution of a well-defined version
of the WDW equation should involve transcendental functions; a property which already appears in the minisuperspace approximation.
As such, the dependence on the cosmological constant
should be in the form of some dimensionless constant $\mathcal{K}$, that is
\beq
A[g_{ij}]=F[\mathcal{K};g_{ij}] \ .
\eeq
Note that $\mathcal{K}$ should be the same for any choice of the time slicing in the ADM foliation, so that $\mathcal{K}$
should be time-independent. Since the Planck length is naturally interpreted as ultraviolet cutoff, we have
\beq
\mathcal{K}=\frac{\cancel{\ell}_P}{L_U} \ ,
\eeq
with $L_U$ a fundamental length describing the geometry of the Universe.
The obvious candidate for $L_U$ is the Hubble radius $R_H=c/H_0=1.36\cdot 10^{26}m$, whose
size is of the same order of the radius of the observable universe and that, besides
$\Lambda$, is the only quantity which is spatially constant.
We then have,
\beq
\mathcal{K}=\frac{\cancel\ell_P}{R_H}=5.96\cdot 10^{-61} \ .
\label{Kappa}\eeq
Furthermore, since $A$ must depend on $\Lambda$, the space-time independence of $\mathcal{K}$ implies that
\beq
A[g_{ij}]=F[D\sqrt\Lambda;g_{ij}] \ ,
\eeq
with $D$, $[D]=L$, a space-time constant.

\noindent Eq.(\ref{Kappa}) would imply that the Planck length is time-dependent. This is
in agreement with the Dirac idea that fundamental constants are dynamical variables.
On the other hand, the most natural candidate for time variation is just the Planck constant
$\hbar$. The point is that the Einstein field equation contains $\Lambda$, $c$ and $G$, and
a possible time dependence of such constants would break diffeomorphism invariance.
Therefore, preserving such an invariance means that only $\hbar$, that in fact appears only in considering the WDW equation, can change. On the other hand,
Eq.(\ref{Kappa}) implies an infrared/ultraviolet duality, where the large scale is given by $R_H$,
whose time dependence is the same of the scale representing the quantum regime, that is (the square root of) $\hbar$.

\noindent
We stress that time variation of fundamental constants is a crucial and widely investigated
subject \cite{Uzan:2002vq}\cite{Uzan:2010pm}\cite{Hart:2017ndk}. In a different context,
time dependence of the Planck constant has been investigated in the interesting paper \cite{Mangano:2015pha}.

\noindent
We conclude by observing that very recently, in \cite{BenAchour:2020xif}, it has been argued by
a different perspective, that the GQHJ theory
introduced in \cite{fm2}, could in fact be at the origin of the cosmological constant.

\section*{Acknowledgements} It is a pleasure to thank N. Bartolo, D. Bertacca, K. Lechner, S. Matarrese, M. Peloso and A. Ricciardone for interesting comments and discussions.

\end{document}